\documentstyle[referee]{l-aa}

\begin{document}

   \thesaurus{06         % A&A Section 6: Form. struct. and evolut. of stars
              (03.11.1;  % Cosmogony,
               16.06.1;  % Planets and satellites: general,
               19.06.1;  % Solar system: general,
               19.37.1;  % Stars: formation of,
               19.53.1;  % Stars: oscillations of,
               19.63.1)} % Stars: structure of.
   \title{Atomic data from the Iron Project}

   \subtitle{XLV. Relativistic transition probabilities for carbon-like
Ar XIII and Fe XXI using Breit-Pauli R-matrix method}
% *   \thanks{Tables of complete data are available
%in electronic form at the CDS via anonymous ftp 130.79.128.5.}

   \author{Sultana N. Nahar }
   \institute{ Department of Astronomy, The Ohio State University,
Columbus, OH 43210, USA }
   \offprints{S.N. Nahar}

\date{Received date; accepted date}

\def\etal{{\it et\thinspace al.}\ }
   \maketitle

\begin{abstract}

The Breit-Pauli R-matrix method developed under the Iron Project 
has been used to obtain extensive sets of oscillator strengths and 
transition probabilities for dipole allowed and intercombination fine 
structure transitions in carbon like ions, Ar XIII and Fe XXI. The complete 
set consists of 1274 fine structure bound energy levels and 198,259 
oscillator strengths for Ar XIII, and 1611 bound levels and 300,079 
oscillator strengths for Fe XXI. These correspond to levels of total angular
momenta of 0 $\leq J \leq 7$ of even and odd parities formed from total
spin of $2S+1$=5,3,1, and orbital angular momenta 0 $\leq L\leq$ 9 with 
$n\leq$ 10, 0 $\leq l\leq 9$ for each ion. The relativistic effects 
are included in the Breit-Pauli approximation. 
The close coupling wavefunction expansion for each ion is represented by 
the lowest 15 fine structure levels of target configuations,
$2s^22p$, $2s2p^2$ and $2p^3$. The energy levels are identified 
spectroscopically using a newly developed identification procedure.
The procedure also makes a correspondence between the fine strucure
energy levels and $LS$ terms. This provides the check for 
completeness of the calculated levels. Comparison is made of the present
energies and the $f$-values with the available observed and theoretical 
values. Present transition probabilitis agree very well with the 
relativistic atomic structure calculations of Mendoza et al. for the 
intercombination transitions, $2s2p^3(^5S^o_2) - 
2s^22p^2(^3P_{1,2},^1D_2)$. This further indicates that the importance 
of the neglected Breit interaction decreases with ion charge and 
constrains the uncertainty in the present calculations to within 
15\% even for the weak transitions.

\end{abstract}

%
%  14.Sep.'90: Demo-Vs.
%________________________________________________________________

\section {Introduction}

With the aim of calculating accurate astrophysical opacities, extensive
set of oscillator strengths ($f$-values) for bound - bound 
transitions and photoionization cross sections for bound - free 
transitions were 
obtained under the Opacity Project (The Opacity Project 1995, 1996; 
Seaton \etal 1994). These data are available through the OP database, the 
TOPbase (Cunto et al. 1993). Determination of opacities requires transition
probabilities or oscillator strengths for all transitions of all ionization 
stages of the constituent elements in the plasma as obtained under the
OP. However, the OP $f$-values were obtained in LS coupling whereas 
transitions among fine structure levels
are often needed in various astrophysical models as well as in experimental
or observational spectral analysis. For example, electron densities in 
solar flares can be determined from the fine structure transitions in Fe
XXI (e.g. Pallavicini et al. 1977). Transition probabilies for fine structure
levels were obtained for a number of Fe ions, such as Fe II (Nahar 1995),
Fe III (Nahar \& Pradhan 1996) and Fe XIII (Nahar 1999) through algebraic
transformation of LS multiplets calculated in close coupling approximation. 
Hence, no relativistic effects were included for these transitions. 

The Breit-Pauli R-matrix (BPRM) (Hummer et al. 1993, Berrington 
\etal 1995) method includes relativistic effects in the 
Breit-Pauli approximation (Scott and Burke 1980, Scott and Taylor 1982). 
It enables the calculations of 
both the dipole allowed ($\Delta S=0$) and the intercombination 
($\Delta S\neq 0$) transitions, in contrast to $LS$ calculations where 
only dipole allowed transitions could be included. Incorporation of the 
relativistic effects in the close coupling R-matrix method yields a 
large number of fine structure transition probabilities with higher 
accuracy.  

However, a major obtstable in the BPRM calculations has been the 
spectroscopic identification of the large number of fine structure energy
levels being calculated. These are obtained as the eigenvalues of the 
BP Hamiltonian labeled only by the total angular momentum and parity, i.e. 
by $J\pi$, which is insufficient for unique identification. 
Except for large scale model calculations, complete identification of
levels is needed for various diagnostics and spectrocopic applications. 
A new procedure is developed to identify these levels by a complete 
set of quantum numbers through analysis of collion channels 
(Nahar \& Pradhan 2000). The procedure also makes a correspondence 
between the fine structure levels and their proper $LS$ terms.

BPRM method has been applied for large scale computations of $f$-values
for a few iron ions, Fe V (Nahar \& Pradhan 2000, Nahar et al 2000),
Fe XXIV and Fe XXV (Nahar and Pradhan 1999). The $f$-values of Fe XXIV
and Fe XXV are found to be very accurate. They agree within a few percent
with the measured and other very accurate calculations available for a 
few transitions. The present work reports oscillator strengths 
for two carbon like ions, Ar XIII and Fe XXI.
Both the dipole allowed and the intercombination 
transitions are considered. The accuray and completeness of the results 
are discussed. While samples of fine structure energy and oscillator
strengths are presented herein, the complete tables will be available 
electronically.

\section{Theory}

In the coupled channel or close coupling (CC) approximation an 
ion is described in terms of an (e + ion) complex that comprises of 
a `target' or core ion, with N bound electrons, and a `free' (N+1)th 
electron that may be either bound or continuum. For a total spin and 
angular symmetry  $SL\pi$ or $J\pi$, of the (N+1) electron system, 
the total wavefunction, $\Psi(E)$, is represented in terms of the target ion states or levels as:

\begin{equation}
\Psi_E(e+ion) = A \sum_{i} \chi_{i}(ion)\theta_{i} + \sum_{j} c_{j}
\Phi_{j},
\end{equation}

\noindent
where $\chi_{i}$ is the target ion wave function in a specific state
$S_iL_i\pi_i$ or level $J_i\pi_i$, and $\theta_{i}$ is the wave function
for the (N+1)th electron in a channel labeled as
$S_iL_i(J_i)\pi_i \ k_{i}^{2}\ell_i(SL\pi) \ [J\pi]$; $k_{i}^{2}$ is the
incident kinetic energy. In the second sum the $\Phi_j$'s are
correlation wavefunctions of the (N+1) electron system that (a) 
compensate for the orthogonality conditions between the continuum and 
the bound orbitals, and (b) represent additional short-range correlation 
that is often of crucial importance in scattering and radiative CC 
calculations for each $SL\pi$.

Details of the theory of Breit-Pauli R-matrix method in the close 
coupling approximation is described in earlier work (Hummer et al. 1993). 
The method yields the solutions of the relativistic CC equations
using the Breit-Pauli Hamiltonian for the (N+1)-electron system to
obtain the total wavefunctions $\Psi_E(e+ion)$. The Breit-Pauli
Hamiltonian is
\begin{equation}
H_{N+1}^{\rm BP}=H_{N+1}+H_{N+1}^{\rm mass} + H_{N+1}^{\rm Dar}
+ H_{N+1}^{\rm so},
\end{equation}
where $H_{N+1}$ is the nonrelativistic Hamiltonian,
\begin{equation}
H_{N+1} = \sum_{i=1}\sp{N+1}\left\{-\nabla_i\sp 2 - \frac{2Z}{r_i}
        + \sum_{j>i}\sp{N+1} \frac{2}{r_{ij}}\right\},
\end{equation}
and the relativistic effects are included through the additional terms 
which are the one-body mass correction term, the
Darwin term and the spin-orbit term respectively. The other relatively
weaker Breit-interaction,
\begin{equation}
H^B = \sum_{i>j}[g_{ij}({\rm so}+{\rm so}')+g_{ij}({\rm ss}')],
\end{equation}
representing the two-body spin-spin and the spin-other-orbit
interactions are not included.

The eigenvalues of the (N + 1)-electron Hamiltonian are the energies
of the states such that,

\begin{equation}
 \begin{array}{l} 
E = k^2 > 0  \longrightarrow continuum~(scattering)~channel (\Psi_F) \\ 
E = - \frac{z^2}{\nu^2} < 0 \longrightarrow bound~state (\Psi_B), 
\end{array}
\end{equation}
where $\nu = n - \mu(\ell)$ is the effective quantum number relative 
to the core level. $\mu(\ell)$ is the quantum defect.

The oscillator strength is proportional to the generalized line 
strength defined, in either length form or velocity form, by the equations
\begin{equation}
S_{\rm L}=
 \left|\left\langle{\mit\Psi}_f
 \vert\sum_{j=1}^{N+1} r_j\vert
 {\mit\Psi}_i\right\rangle\right|^2 \label{eq:SLe}
\end{equation}
and
\begin{equation}
S_{\rm V}=\omega^{-2}
 \left|\left\langle{\mit\Psi}_f
 \vert\sum_{j=1}^{N+1} \frac{\partial}{\partial r_j}\vert
 {\mit\Psi}_i\right\rangle\right|^2. \label{eq:SVe}
\end{equation}
In these equations $\omega$ is the incident photon energy
in Rydberg units, and $\mit\Psi_i$ and $\mit\Psi_f$ are the wave
functions representing the initial and final states respectively. 

Using the energy difference, $E_{ji}$, between the initial and final
states, the oscillator strength, $f_{ij}$, for the transition can be
obtained from $S$ as

\begin{equation}
f_{ij} = {E_{ji}\over {3g_i}}S,
\end{equation}

\noindent
and the Einstein's A-coefficient, $A_{ji}$, as

\begin{equation}
A_{ji}(a.u.) = {1\over 2}\alpha^3{g_i\over g_j}E_{ji}^2f_{ij},
\end{equation}

\noindent
where $\alpha$ is the fine structure constant, and $g_i=2J_i+1$, 
$g_j=2J_j+1$ are the statistical weight factors of the initial and final 
states, respectively. In terms of c.g.s. unit of time,
\begin{equation}
A_{ji}(s^{-1}) = {A_{ji}(a.u.)\over \tau_0},
\end{equation}

\noindent
where $\tau_0 = 2.42^{-17}$s is the atomic unit of time. The lifetime 
of a level can be obtained from the A-values of the level as, 
\begin{equation}
\tau_f = {1\over A_f},
\end{equation}
where $A_f$ is the total radiative transition probability for the 
level f, i.e.,
\begin{equation}
A_f = {\sum_i A_{fi}}.
\end{equation}

\section{Computations}

The R-matrix calculations begin with the target wavefunction obtained
through configuration interaction atomic structure calculation.
Present target wavefunctions are obtained from the atomic structure 
calculations using SUPERSTRUCTURE (Eissner et al 1974).
The wavefunctions of Ar XIII and Fe XXI are represented by expansions 
of 15 fine structure levels of the target or the core ion, Ar XIV and 
Fe XXII, belonging to the 8 lowest $LS$ terms: $2s^22p(^2P^o)$, 
$2s2p^2(^4P,^2P,^2D,^2S)$, and $2p^3(^4S^o,^2D^o,^2P^o)$. The corresponding
15 fine structure levels alongwith their energies are listed in Table 1.  
The energies are the observed ones (Kelly, NIST, for Ar XIV, and Sugar and 
Corliss 1985 for Fe XXII) for improved accuray. The correlation and 
spectroscopic configurations and the values of the scaling parameter 
($\lambda$) in the Thomas-Fermi potential for each orbital of 
Ar XIV and Fe XXII in the atomic structure calculations are also
listed in Table 1.  The bound-channel term, the second expansion in 
the wavefunction, Eq.(1), includes all possible configurations upto 
$2p^4$, $3s^2$, $3p^2$, and $3d^2$ for both Ar XIII and Fe XXI.

The one- and two-electron radial integrals are computed by STG1 of the 
BPRM codes using the one-electron target orbitals generated by 
SUPERSTRUCTURE. The number of continuum R-matrix basis functions  
is 12 for each ion. The calculations consider all possible bound levels 
for 0$\leq J\leq$8 with $n < 10, \ \ell \leq n-1$, $0\leq L \leq$ 9, and
$(2S+1)$=1,3,5, even and odd parities. The intermediate coupling 
calculations are carried out on recoupling the $LS$ symmetries in a 
pair-coupling representation in stage RECUPD. The (e + core) Hamiltonian 
matrix is diagonalized for each resulting $J\pi$ in STGH. 

STGB of the BPRM codes calculates the fine structure energy levels and 
their wavefunctions. As fine structure causes large number
of closely spaced energy levels, STGB requires to use an energy mesh of
effective quantum number, $\Delta \nu$=0.001, an order of magnitude
finer than needed in the LS coulping case to avoid levels missing. 
This increases the computation time considerably. 

The oscillator strengths ($f$-values) are computed using STGBB of the
BPRM codes. STGBB computes the transition matrix elements using the
bound wavefuncitons created by STGB and angular algebra for the dipole
moment calculated by STGH. The STGBB computations are also considerably
CPU time extensive due to large number of dipole allowed and
intercombination transitions among the fine structure levels.

The BPRM method, which uses the collision theory, describes the energy 
levels with channel identification for a given total $J\pi$. 
A level identification procedure as described in Nahar and Pradhan 
(2000) is implemented to identify the levels of Ar XIII and Fe XXI. 
The identification scheme is based on the analysis of quantum defects 
and channel wavefunctions similar to that under the Opacity Project 
(Seaton 1987).
The levels are designated with  possible identification of 
$C_t(S_tL_t\pi_t)J_tnlJ(SL)\pi$ where $C_t$, $S_tL_t\pi_t$, $J_t$ are
the configuration, $LS$ term and parity, and total angular momentum of 
the core or target, $nl$ are the principle and orbital quantum numbers 
of the outer or the valence electron, $J$ and $SL\pi$ are the total 
angular momentum, possible $LS$ term and parity of the 
$(N+1)$-electron system. Each level is associated with a numer of
collision channels. The identification scheme carries out the
quantum defect analysis of the contributing channels with maximum
channel percentage weight, i.e., the dominant channels that determine 
the proper configurations and terms of the core 
and the outer electrons. 

The principle quantum number, $n$, of the outer
electron of a level is determined from its $\nu$, and a Rydberg series
of levels is identified from the effective quantum number. Hence,
$\nu$ of the lowest member (level with
the lowest principal quantum number of the valence electron) of
a Rydberg series is determined from quantum defect analysis of all
the computed levels for each partial wave $l$. The lowest partial wave
has the highest quantum defect. A check is maintained to differentiate
the quantum defect of a $'s'$ electron with that of an equivalent electron
state which has typically a large value in the close coupling calculations.

Two levels with the same configuration and set of quantum numbers can 
actually be two independent levels due to outer electron spin 
addition/ subtraction to/from
the parent spin angular momentum, i.e. $S_t\pm s = S$. 
The lower energies are normally
assigned with the higher spin multiplicity. However, the energies and
effective quantum numbers ($\nu$) of levels of higher and lower spin
multiplicity  can be very close to each other, in which case the spin
multiplicity assignment may be uncertain.

Following level identification, a direct correspondence is made with 
standard spectroscopic designations that
follow different coupling schemes, such as between $LS$ and $JJ$.
he correspondence provides
the check for completeness of calculated set of levels or the levels
missing. The level identification procedure involves considerable
manipulation of the bound level data and, although it has been encoded
for general applications, still requires analysis and interpretation of
problem cases of highly mixed levels that are difficult to identify.
These are processed for possible correspondence
between the sets of $SL\pi$ and $J\pi$ of the same configuration, and
are checked for the completeness of all fine structure levels within 
an $LS$ term. A computer program, PRCBPID, is developed to carry out the 
identification scheme (Nahar and Pradhan 2000).

\section {Results and Discussions}

The fine structure energy levels and oscillator strengths for the
dipole allowed and the intercombination transitions are discussed
in separate subsections below.

\subsection{Energy Levels}

A total of 1274 bound fine structure energy levels of Ar XIII and 1611 
levels of Fe XXI are obtained. These correspond to total angular momenta
of 0 $\leq J \leq$ of both even and odd parities formed from spin
multiplicities of $2S+1$ = 5,3,1, total orbital angular momenta of
0 $\leq L \leq$ 9 with $n\leq$ 10, 0$\leq l \leq (n-1)$.

The energy levels have been identified by an identification procedure as 
mentioned above. These levels are presented in two formats: 
(i) in $J\pi$ order for practical applications and (ii) in $LS$ term order for spectroscopy and
completeness check up.

Table 2a presents a partial set of energy levels of Ar XIII in format
(i), i.e., in $J\pi$ order (the complete table is available 
electronically). At the top of each set. $N_J$ is the total number of 
energy levels for symmetry of $J\pi$. For example, there are 56 fine
structure levels of Ar XIII with $J\pi$ = $0^e$. However, in the table, 
only part of the levels belonging to symmetry, $J\pi$, are presented 
for illustrations. For each level with a valence electron, the 
effective quantum number, $\nu~=~z/\sqrt(E-E_t)$ where $E_t$ is the
immediate target threshold, is given next to the energies. $\nu$ is not
defined for equivalent electron levels and is not given for them. Each 
level is assigned to one or more 
$LS$ terms. If number of possible term is more than one, all are specified.
Initially the term designation for a level with multiple $LS$ 
term assignment can be made using Hund's rule, that the 
term with higher angular momentum lies lower in energy.
For example, the three levels 4, 5, and 6 of set $J\pi$ = 1$^e$ are
assigned with 3 possible terms, $^3(SPD)^e$, whereupon 
the first level, i.e., level 4 is $^3D^e$,
while level 5 is $^3P^e$, and level 6 is $^3S^e$. One reason for specifying
all possible terms is that the order of calculated energy levels may 
not match exactly that of the measured ones. The other reason is that 
Hund's rule may not apply to all cases for complex ions; nonetheless
it is useful to establish completeness. Similar sets of
levels for Fe XXI are presented in Table 3a. 

Table 2b presents the levels in  ascending energy order regardless 
of $J\pi$ values, and are grouped together according to the
same configuration to show the correspondence between two sets of
representations, in $J$-levels and in $LS$ terms. (Listing of the 
lowest levels of equivalent electron states are omitted as they 
are given in energy comparison table).  The grouping of levels  
provides the check for completeness of sets of energy levels that 
should belong to the corresponding $LS$ term, and
detects any missing level. The title line of each set of levels 
in Table 2b lists spin multiplicity ($2S$+1), parity, all possible 
$L$-values that can be formed from the core or target term, and outer 
or the valence electron angular momentum. The $J$-values for each 
possible $LS$ term are
specified within the parentheses next to the value of $L$. 'Nlv'
is the total number of $J$-levels that are expected from this set of 
$LS$ terms. This line is followed by the set of BPRM energy levels of 
same configurations. 'Nlv(c)', at the end of the set, specifies the total 
number of $J$-levels obtained. If Nlv = Nlv(c) for a set of levels, 
the calculated energy set is complete. The correspondence of couplings
and completeness of levels are carried out by the program PRCBPID 
which also detects and prints the missing levels. For example,
in the set for $^3(G,H,I)^e$ for Ar XIII near the end of Table 2b, 
the set is found to be incomplete where four levels of $J$=5,4,6,5 are 
missing. Sets with missing levels usually lie in the high energy region.
Each level of a set is further identified by all possible $LS$ terms 
(specified in the last column of the set). The multiple $LS$ terms 
can be reduced to the most probable (but approximately) one using 
Hund's rule, as explained 
above. It may be noted that levels are grouped consistently in
closely spaced energies and in effective quantum numbers
confirming the designation of the $LS$ terms. Similar sets of levels of 
Fe XXI are presented in Table 3b.

The BPRM energy levels for Ar XIII and Fe XXI are compared in Table 4
with the limited number of levels observed. The 55 observed levels 
of Ar XIII (Kelly, NIST) are not in the compiled list by the NIST. 
However, the calculated fine structure energies agree with these 
observed ones to about 1\% for most of the levels. The difference is 
upto 9\%. The agreement 
between the observed and calculated energis is much better for Fe XXI
(Table 4). There are 39 observed energy levels of Fe XXI (Sugar and 
Corliss 1985). Present energies agree with the observed ones to less 
than or about 1\% for all except two levels that agree within 2\%. 
In Table 4 the column next to the $J$-column is for the 
energy level index, $I_J$. The energy level index specifies the position,
in ascending order, of the level in the calculated set of energies of
$J\pi$ symmetry. It is necessary to use the level indices to make 
correspondence between the calculated and observed levels for later use.

\subsection{Oscillators strengths}

The total number of oscillator strengths ($f$-values) obtained for 
fine structure bound-bound transitions in Ar XIII is 198,259 and 
for Fe XXI is 300,079. These include both the dipole
allowed ($\Delta S$ =0) and the intercombination ($\Delta S \neq$ 0) 
transitions. Complete sets of $f$-values for both Ar XIII and Fe XXI 
are available electronically.

Partial sets of the oscillator strengths for Ar XIII and Fe XXI  
are presented in Tables 5 and 6 respectively. The format of the tables
is the same as that of the electronic files for the $f$-values. At the 
top of each table, the two numbers are the nuclear charge (Z = 18 for 
Ar XIII and = 26 for Fe XXI) and number of electrons in the 
ion, $N_{elc}$ (= 6 for carbon like ions). Below this line are the
sets of oscillator strengths belonging to a pair of symmetries, 
$J_i\pi_i~-~J_j\pi_j$, specified at the top. The symmetries are expressed
in the form of $2J_i$ and $\pi_i$ (=0 for even and =1 for odd parity), 
$2J_j$ and $\pi_j$. For example, Tables 5 and 6 present partial 
transitions among the levels of symmetries $J=0^e$ and $J=1^o$ for
Ar XIII and Fe XXI respectively. The line following 
the transition symmetries are the number of bound levels, $N_{Ji}$ 
and $N_{Jj}$. This line is followed by $N_{Ji}\times N_{Jj}$ number of
transitions. The first two columns are the level indices, $I_i$ and 
$I_j$ (as mentioned above) for the energy indices of the levels, and
the third and the fourth columns are their energies, $E_i$ and $E_j$, 
in Rydberg unit. The fifth column is the $gf_L$ for the allowed 
transitions ($\Delta J$ = 0,$\pm 1$). $f_L$ is the oscillator strength
in length form, and $g~=~2J+1$ is the statistical weight factor of 
the initial or the lower level. A negative value for $gf$ means that 
$i$ is the lower level, while a positive one means that $j$ is the 
lower level. Column six is the line strength (S). The last column 
in the table gives the transition probability, $A_{ji}(sec^{-1})$. 
Complete spectroscopic identification of the transition can be 
obtained from Tables 2a and 3a by referrring to the values of $J_i\pi_i$, 
$I_i$, $J_j\pi_j$, and $I_j$. For example, 
the first transition for Fe XXI in Table 6 corresponds to, as identified
in Table 3a, dipole allowed transition $2s^22p^2(^3P^e_0)(I_i=1) 
\rightarrow 2s2p^3(^3D^o_1)(I_j=1)$.

A set of transition probabilities for both Ar XIII and Fe XXI has been
reprocessed such that observed energy differences, rather than the
calculated ones, are used to calculate the $f$- and $A$-values from BPRM
line strengths, $S$-values (Eqs. 7 and 8). This improves the accuracy of 
the $f$- and $A$-values of the relevant transitions; observed 
energies have lower uncertainties than the calculated ones. Hence, for
any comparison or spectral diagnostics, values from these tables
should be used. The reprocessing of $f$- and $A$-values has been carried 
out for the transitions among all the observed levels. 
The reprocessed set of oscillator strengths consists of 333 transitions
of Ar XII and 184 of Fe XXI (the reprocssed sets are also available
electronically). Sample sets of the reprocessed oscillator strengths 
are presented for both the ions in Tables 7a and 8a. Transitions are
listed in $J\pi$ order, and in contrast to the large complete files 
(Tables 5 and 6) they are given with complete identification. 
The level index, $I_i$, for each energy level in the tables is given 
next to the $J$-value for an easy link to the complete $f$-file. 

In a model calculation requring large number of transitions, the 
reprocessed values of $f$ and $A$ for Ar XIII and Fe XXI should replace 
those in the large complete files. For example, the transition 
$J=0^e(I_i=1)\rightarrow J=1^o(I_j=1)$ in Table 7a corresponds to the 
first transition in Table 5, and hence, the $f$- and $A$-values of Table 
7a should replace those in Table 5. The overall replacement in Tables 
5 or 6 can be carried out by sorting out the reprocessed transitions 
through their energy level indices. The set of level indices of each
symmetry $J\pi$ for both Ar XIII and Fe XXI that correspond to these 
transitions are given in Table 9. All allowed transitions 
($\Delta J=0,\pm 1$ except 
among $J$=0 levels) among these levels have been considered. 

The reprocessed transitions are further ordered in terms of their 
configurations along with the $LS$ terms. This enables one to obtain the 
$f$-values for the $LS$ multiplets and check the completeness of the 
set of fine structure components belonging to the multiplet. However, 
the completeness depends also on the observed set of fine structure 
levels since reprocessed transitions correspond only to those levels 
that have been observed. A partial set of these transitions is 
presented in Table 7b for Ar XIII and Table 8b for Fe XXI (the 
complete table is available electronically). The $LS$ 
multiplets are useful for various comparisons with other calculations 
and experiment where fine structure transitions can not be resolved. 
The oscillator strengths are also compared with available ones in 
Tables 7b and 8b.

The oscillator strengths for a large number of fine structure 
transitions in Ar XIII and other carbon like ions have been obtained by 
Zhang and Sampson (1997) 
in the relativistic distorted wave calculations. They consider the 
transitions among the n = 2 and n = 3 levels. The other source of
transition probabilities for Ar XIII is TOPbase (the database for
the Opacity Project data) where transition probabilties for
$LS$ multiplets are given for states with $n \leq 10$. These values were 
calculated by Luo and Pradhan (1989) in the close coupling approximation
using non-relativistic R-matrix method.  
Comparison of the present BPRM oscillator strengths with the previous
calculations in Table 7b shows that present $LS$ multiplets obtained from 
the fine structure components agree quite well with those of Luo and
Pradhan except for the transition, $2s2p^3(^3D^o)~-~2p^4(^3P)$. However,
comparison with Zhang and Sampson shows various degrees of agreement. 
There is good agreement for some fine structure components compared 
to others in the same $LS$ multiplet, e.g. for 
$2s^22p^2(^3P)~-~2s2p^3(^3D^o)$. The largest difference is found in the weak 
transitions. 

The $f$-values of Fe XXI have been compiled by Fuhr et al. (1988) and 
Shirai et al. (1990) (both references have the same values). 
Similar to the above case of Ar XIII, the Fe XXI oscillator strengths  
compare well for some but poorly for other fine structure components 
within a multiplet, with previous calculations (Table 8b). Similarly 
the multiconfiguration Dirac-Fock calculations by Cheng et al (1979) 
compare well with the BPRM $f$-values for some components, but poorly 
with others. The $LS$ multiplets are compared with those of Luo and 
Pradhan (1989) and in general agree very well with the present BPRM 
$A$-values. 

Recently Mendoza et al. (1999) have calculated the transition
probabilties for the intercombination transitions 
$2s2p^3(^5S^o_2)~-~2s^22p^2(^3P_{1,2},^1D_2)$ in the carbon isoelectronic 
sequence ions. Through extensive relativistic atomic structure 
calculations they study the effects of Breit interactions in the
transition probabilties for these weak transitions. They estimate an 
accuracy better than 10\% for A-values belonging to the ground 
levelsi $^3P_{1,2}$ and 20\% for the $^1D_2$. Their values for Ar XIII agree
very well with the present BPRM $A$-values (Table 7b), especially those
to the ground levels. The agreement is 4\% for the $^5S^o_2~-~^3P_1$
and 0.01\% for the $^5S^o_2~-~^3P_2$ transitions. The agreement is 
14\% for the $^5S^o_2~-~^1D_2$ for which they assign a larger uncertainty.
For Fe XXI, present $f$-values for the transitions 
$2s2p^3(^5S^o)~-~2s^22p^2(^3P_{1,2}, ^1D_2)$ agree very well with those 
by Cheng et al. The agreement is also quite good with the $A$-values 
obtained in elaborate calculations by Mendoza et al. (1999); about 
4\% for both the transitions to the ground term, $^5S^o_2~-~^3P_{1,2}$, 
and 5\% for the $^5S^o_2~-~^1D_2$ transition.

Comparison of the present transition probabilities for the intercombination
transitions with those of Mendoza et al (1999) provides a measure of 
uncertainty of the present BPRM results. They studied the effect of Breit
spin-spin and spin-other orbit interactions which
are not included in the present work. The good agreement with
them for these very weak transtions indicate the uncertainty of the present
results to a maximum of 14\% for most of the transitions. 
The discrepancy between the oscillator strengths in length and velocity 
forms ($f_L$ and $f_V$) is an indicator of accuracy consistent with the 
method of calculations. In the present results, it is found that the 
dispersion of the $f_L$ and $f_V$ values for Fe XXI is mainly in the very
weak transitons. However, the dispersion is larger for Ar XIII; it is
expected that the dispersion is caused mainly by the values of $f_V$. 
Inclusion of more configurations in the Ar XIII wavefunction expansion 
could have improved the agreement. However, the length form in a close
coupling approximation is more accurate than the velocity
form as it depends more on the asymptotic form of the wavefunction is
better represented in the R-matrix method.

\section {Conclusion}

We have described some of the first large-scale ab initio relativistic
calculations of transition probabilities now in progress under the Iron
Project. Large datasets for the fine structure energy levels and 
oscillator strengths for the carbon like Ar XIII and Fe XXI are presented.
These exceed far in the presently available data for these two ions. 
All the calculated energy levels, 1274 for Ar XIII and 1611 for Fe XXI,
have been spectroscopically identified and checked for completenss. All 
data are available electronically. Part of the $f$-values have been 
reprocessed using available observed energies for better accuracy.

\begin{acknowledgements}
This work was partially supported by U.S. National Science Foundation 
(AST-9870089) and NASA (NAG5-7903).  The computational work was
carried out on the Cray T94 at the Ohio Supercomputer Center in 
Columbus, Ohio.

\end{acknowledgements}

%\newpage

%___________________________________ Two column table (place early!)

\pagebreak

\begin{table}
\noindent{Table 1. The 15 fine strucuture levels of the core (target)
ions Ar XIV and Fe XXII in the close coupling eigenfunction expansion 
of Ar XIII and Fe XXI. 
List of configurations, and values of the scaling parameter ($\lambda$) 
for each orbital are given below the table.
  \\ }
%\small
\scriptsize
\begin{tabular}{llrrllrrllrrllrr}
\noalign{\smallskip}
\hline
\noalign{\smallskip}
& Term & $J_t$ & $E_t$(Ry) & & Term & $J_t$ & $E_t$(Ry) &
& Term & $J_t$ & $E_t$(Ry) & & Term & $J_t$ & $E_t$(Ry) \\
 \noalign{\smallskip}
 \hline
 \noalign{\smallskip}
 \multicolumn{8}{c}{Ar XIV } & \multicolumn{8}{c}{Fe XXI } \\
 \noalign{\smallskip}
 \hline
 \noalign{\smallskip}
1 & $2s^22p(^2P^o)$ & 1/2 & 0.0 & 
9 & $2s2p^2(^2P^e)$ & 1/2 & 4.9685 &
1 & $2s^22p(^2P^o)$ & 1/2 & 0.0 &
9 & $2s2p^2(^2S^e)$ & 1/2 & 8.91420 \\
2 & $2s^22p(^2P^o)$ & 3/2 & 0.20647 &
10& $2s2p^2(^2P^e)$ & 3/2 & 5.0544 & 
2 & $2s^22p(^2P^o)$ & 3/2 & 1.07776 &
10& $2s2p^2(^2P^e)$ & 3/2 & 9.04241 \\
3 & $2s2p^2(^4P^e)$ & 1/2 & 1.7925& 
11& $2p^3(^4S^o)$   & 3/2 & 6.2455 & 
3 & $2s2p^2(^4P^e)$ & 1/2 & 3.68653 &
11& $2p^3(^4S^o)$  & 3/2 & 11.44278 \\
4 & $2s2p^2(^4P^e)$ & 3/2 & 1.8709 & 
12& $2p^3(^2D^o)$   & 3/2 & 7.1115 & 
4 & $2s2p^2(^4P^e)$ & 3/2 & 4.19365 &
12& $2p^3(^2D^o)$   & 3/2 & 12.72512 \\
5 & $2s2p^2(^4P^e)$ & 5/2 & 1.9756& 
13& $2p^3(^2D^o)$   & 5/2 & 7.1361 &
5 & $2s2p^2(^4P^e)$ & 5/2 & 4.67717 &
13& $2p^3(^2D^o)$ & 5/2 & 13.00269 \\
6 & $2s2p^2(^2D^e)$ & 3/2 & 3.7387 & 
14& $2p^3(^2P^o)$   & 1/2 & 8.0301 & 
6 & $2s2p^2(^2D^e)$ & 3/2 & 6.71166 &
14& $2p^3(^2P^o)$   & 1/2 & 14.30352 \\
\hline
\end{tabular}
\end{table}

\pagebreak

\begin{table}
\noindent{Table 2a. Identified fine strucuture energy levels of Ar XIII. 
$N_J$=total number of levels for the symmetry $J\pi$. $SL\pi$ lists 
possible set of $LS$ terms. 
  \\ }
%\small
\scriptsize
\begin{tabular}{rllrrrcrl}
\noalign{\smallskip}
\hline
\noalign{\smallskip}
 i & $C_t$ & $S_tL_t\pi_t$ & $J_t$ & $nl$ & $J$ & $E(Ry)$ & $\nu$ & $SL\pi$ \\
 \noalign{\smallskip}
 \hline
 \noalign{\smallskip}
 \multicolumn{9}{c}{$N_J$= 56,~~~~~$J\pi$= $0^e$        } \\
 \noalign{\smallskip}
 \hline
 \noalign{\smallskip}
   1 & $2s22p2 $ &         &      & $   $ &  0 & -5.04159E+01 &       &
$^3P e $ \\
   2 & $2s22p2 $ &         &      & $   $ &  0 & -4.89054E+01 &       &
$^1S e $ \\
   3 & $2p4    $ &         &      & $   $ &  0 & -4.14331E+01 &       &
$^3P e $ \\
   4 & $2p4    $ &         &      & $   $ &  0 & -3.96913E+01 &       &
$^1S e $ \\
   5 & $2s22p  $ & $^2P^o$ &  1/2 & $ 3p$ &  0 & -2.05345E+01 &  2.87 &
$^3P e $ \\
   6 & $2s22p  $ & $^2P^o$ &  3/2 & $ 3p$ &  0 & -1.97019E+01 &  2.91 &
$^1S e $ \\
\hline
\end{tabular}
\end{table}

\pagebreak

\begin{table}
\noindent{Table 2b. Ordered and identified fine strucuture energy levels 
of Ar XIII. Nlv=total number of levels expected for the possible $LS$ 
terms listed and Nlv(c) = number of levels calculated. $SL\pi$ lists 
the possible $LS$ terms for each level. (See texts for details.) 
 \\ }
%\small
\scriptsize
\begin{tabular}{lcrrrcrl}
\noalign{\smallskip}
\hline
\noalign{\smallskip}
$C_t$ & $S_tL_t\pi_t$ & $J_t$ & $nl$ & $J$ & E(cal) & $\nu$ & $SL\pi$ \\
 \noalign{\smallskip}
 \hline
 \noalign{\smallskip}
 \multicolumn{8}{l}{ Eqv electron/unidentified levels, parity: e
         }\\
 2p4      &     &    &   &  2& -4.16174E+01 &  0.00 &  3~ P e       \\
 2p4      &     &    &   &  1& -4.14770E+01 &  0.00 &  3~ P e       \\
 2p4      &     &    &   &  0& -4.14331E+01 &  0.00 &  3~ P e       \\
 \noalign{\smallskip}
 \multicolumn{8}{l}{Nlv(c)=~ 3~ : set complete
                                                                       }
\\
 \noalign{\smallskip}
 \hline
 \noalign{\smallskip}
 \multicolumn{1}{l}{Nlv=~ 3,~ 3,o:} &
 \multicolumn{7}{l}{ P ( 2 1 0 )
             } \\
 \noalign{\smallskip}
 2s22p    &(2Po)& 1/2& 3s&  0& -2.20575E+01 &  2.77 &  3~ P o       \\
\hline
\end{tabular}
\end{table}

\pagebreak

\begin{table}
\noindent{Table 3a. Identified fine strucuture energy levels of Fe XXI. 
$N_J$=total number of levels for the symmetry $J\pi$.  $SL\pi$ lists 
possible set of $LS$ terms.
  \\ }
%\small
\scriptsize
\begin{tabular}{rllrrrcrl}
\noalign{\smallskip}
\hline
\noalign{\smallskip}
 i & $C_t$ & $S_tL_t\pi_t$ & $J_t$ & $nl$ & $J$ & $E(Ry)$ & $\nu$ & $SL\pi$ \\
 \noalign{\smallskip}
 \hline
 \noalign{\smallskip}
 \multicolumn{9}{c}{$N_J$= 64,~~~~~$J\pi$= $0^e$        } \\
 \noalign{\smallskip}
 \hline
 \noalign{\smallskip}
   1 & $2s22p2 $ &         &      & $   $ &  0 & -1.24292E+02 &       &
$^3P e $ \\
   2 & $2s22p2 $ &         &      & $   $ &  0 & -1.20849E+02 &       &
$^1S e $ \\
   3 & $2p4    $ &         &      & $   $ &  0 & -1.08406E+02 &       &
$^3P e $ \\
   4 & $2p4    $ &         &      & $   $ &  0 & -1.05511E+02 &       &
$^1S e $ \\
   5 & $2s22p  $ & $^2P^o$ &  1/2 & $ 3p$ &  0 & -5.21862E+01 &  2.91 &
$^3P e $ \\
   6 & $2s2p2  $ & $^4P^e$ &  1/2 & $ 3s$ &  0 & -5.03119E+01 &  2.86 &
$^3P e $ \\
\hline
\end{tabular}
\end{table}
%\clearpage

\pagebreak

\begin{table}
\noindent{Table 3b. Ordered and identified fine strucuture energy 
levels of Fe XXI. Nlv=total number of levels expected
for the possible $LS$ terms listed and Nlv(c) = number of levels
calculated. $SL\pi$ lists the possible $LS$ terms for each level. (See
texts for details.) 
 \\ }
%\small
\scriptsize
\begin{tabular}{lcrrrcrl}
\noalign{\smallskip}
\hline
\noalign{\smallskip}
$C_t$ & $S_tL_t\pi_t$ & $J_t$ & $nl$ & $J$ & E(cal) & $\nu$ & $SL\pi$ \\
 \noalign{\smallskip}
 \hline
 \noalign{\smallskip}
 \multicolumn{8}{l}{Nlv(c)=~ 1~ : set complete
                                                                       }
\\
 \noalign{\smallskip}
 \hline
 \noalign{\smallskip}
 \multicolumn{1}{l}{Nlv=~ 3,~ 3,o:} &
 \multicolumn{7}{l}{ P ( 2 1 0 )
             } \\
 \noalign{\smallskip}
 2s22p    &(2Po)& 1/2& 3s&  0& -5.48009E+01 &  2.84 &  3~ P o       \\
 2s22p    &(2Po)& 3/2& 3s&  1& -5.45579E+01 &  2.81 &  3~ P o       \\
 2s22p    &(2Po)& 3/2& 3s&  2& -5.37314E+01 &  2.84 &  3~ P o       \\
 \noalign{\smallskip}
 \multicolumn{8}{l}{Nlv(c)=~ 3~ : set complete
                                                                       }
\\
 \noalign{\smallskip}
 \hline
 \noalign{\smallskip}
 \multicolumn{1}{l}{Nlv=~ 1,~ 1,o:} &
 \multicolumn{7}{l}{ P ( 1 )
             } \\
\hline
\end{tabular}
\end{table}

\begin{table}
\noindent{Table 4. Comparison of calculated BPRM energies, $E_c$, with 
the observed ones (Sugar and Corliss 1985), $E_o$, Ar XIII and Fe XXI. 
$I_J$ is the level index for the energy position in symmetry $J\pi$. The
asterisk next to the $J$-value indicates that the term has incomplete set 
of observed fine structure levels. \\
}
%\label{TabSecInst}
\scriptsize
\begin{tabular}{llrrrrlllrrr}
\hline
\noalign{\smallskip}
\multicolumn{2}{c}{Level} & $J$ & $I_J$ & \multicolumn{1}{c}{$E_c$(Ry)} &
\multicolumn{1}{c}{$E_o$(Ry)} & \multicolumn{2}{c}{Level} & $J$ & $I_J$
& \multicolumn{1}{c}{$E_c$(Ry)} & \multicolumn{1}{c}{$E_o$(Ry)} \\
            \noalign{\smallskip}
\hline
            \noalign{\smallskip}
\multicolumn{12}{c}{Ar XIII} \\
            \noalign{\smallskip}
\hline
            \noalign{\smallskip}
$2s^22p^2   $ &   $^3P$   &  2 & 1 & 50.133 &   50.209  & 
$2s^22p3d   $ &   $^3P^o$ &  2 & 9 & 19.053 &   19.149 \\ 
$2s^22p^2   $ &   $^3P$   &  1 & 1 & 50.243 &   50.326  & 
$2s^22p3d   $ &   $^3P^o$ &  1 & 8 & 19.037 &   19.112 \\ 
$2s^22p^2   $ &   $^3P$   &  0 & 1 & 50.332 &   50.416  & 
$2s^22p3d   $ &   $^3P^o$ &  0 & 3 & 19.025 &   19.079 \\ 
$2s^22p^2   $ &   $^1D$   &  2 & 2 & 49.558 &   49.621  & 
$2s^22p3d   $ &   $^1P^o$ &  1 & 9 & 18.740 &   18.772 \\
$2s^22p^2   $ &   $^1S$   &  0 & 2 & 48.856 &   48.905  & 
$2s^22p3d   $ &   $^1F^o$ &  3 & 4 & 18.736 &   18.787 \\ 
$2s2p^3     $ &   $^5S^o$ &  2 & 1 & 48.274 &   48.346  & 
$2s2p^23p    $ &   $^3D^o$ &  3 & 8 & 17.126 &   16.822 \\
\hline
\end{tabular}
\end{table}

\pagebreak

\begin{table}
\noindent{Table 5. Oscillator strengths and transition probabilities for 
Ar XIII (see text for explanation).  \\ }
%\small
\scriptsize
\begin{tabular}{lrccccc}
\noalign{\smallskip}
\hline
\noalign{\smallskip}
\multicolumn{7}{l}{~~~18~~~~6} \\
\multicolumn{7}{l}{~~~~0~~~~0~~~~~~~~~2~~~~1} \\
56  & 136 & $E_i(Ry)$ & $E_j(Ry)$ & $gf_L$ & $ S  $ & $A_{ji}(sec^{-1})$ \\
 \noalign{\smallskip}
    1 &  1&-5.04159E+01&-4.65272E+01&-7.432E-02& 5.734E-02& 3.009E+09\\
    1 &  2&-5.04159E+01&-4.58436E+01&-3.719E-02& 2.440E-02& 2.082E+09\\
    1 &  3&-5.04159E+01&-4.45972E+01&-7.896E-02& 4.071E-02& 7.158E+09\\
    1 &  4&-5.04159E+01&-4.39490E+01&-8.538E-03& 3.961E-03& 9.561E+08\\
    1 &  5&-5.04159E+01&-2.20006E+01&-5.614E-02& 5.927E-03& 1.214E+11\\
\hline
\end{tabular}
\end{table}

\begin{table}
\noindent{Table 6. Oscillator strengths and transition probabilities 
for Fe XXI (see text for explanation).  \\ }
%\small
\scriptsize
\begin{tabular}{lrccccc}
\noalign{\smallskip}
\hline
\noalign{\smallskip}
\multicolumn{7}{l}{~~~26~~~~6} \\
\multicolumn{7}{l}{~~~~0~~~~0~~~~~~~~~2~~~~1} \\
64  & 157 & $E_i(Ry)$ & $E_j(Ry)$ & $gf_L$ & $ S  $ & $A_{ji}(sec^{-1})$ \\
 \noalign{\smallskip}
    1 &  1&-1.24292E+02&-1.17168E+02&-8.556E-02& 3.603E-02& 1.163E+10\\
    1 &  2&-1.24292E+02&-1.15815E+02&-2.267E-02& 8.023E-03& 4.362E+09\\
    1 &  3&-1.24292E+02&-1.14241E+02&-3.606E-02& 1.076E-02& 9.752E+09\\
    1 &  4&-1.24292E+02&-1.12695E+02&-1.038E-04& 2.685E-05& 3.738E+07\\
    1 &  5&-1.24292E+02&-5.45579E+01&-5.459E-02& 2.348E-03& 7.108E+11\\
\hline
\end{tabular}
\end{table}

\pagebreak

\begin{table}
\noindent{Table 7a. Sample set of reprocessed $f$- and $A$-values for
Ar XIII where observed transition energies are replaced by the 
calculated ones. $I_i$ and $I_j$ are the level indices. 
\\ }
\scriptsize
\begin{tabular}{llllccrrcc}
\noalign{\smallskip}
\hline
\noalign{\smallskip}
\multicolumn{1}{c}{$C_i$} &\multicolumn{1}{c}{$C_j$} & 
\multicolumn{1}{c}{$S_iL_i\pi_i$} & \multicolumn{1}{c}{$S_jL_j\pi_j$} & 
$g_i(I_i)$ & $g_j(I_j)$ & $E_i(Ry)$ & $E_j(Ry)$ 
& $f$ & $A(sec^{-1})$ \\
 \noalign{\smallskip}
\hline
 \noalign{\smallskip}
 $2s^22p^2      $ & $-~2s2p^3       $ & $^3P^e$ & $^3D^o$ &  1 (  1)&  3
(  1)& 50.3320 &  46.4760 &  7.370E-02 &  2.93E+09 \\
 $2s^22p^2      $ & $-~2s2p^3       $ & $^3P^e$ & $^3P^o$ &  1 (  1)&  3
(  2)& 50.3320 &  45.8140 &  3.675E-02 &  2.01E+09 \\
 $2s^22p^2      $ & $-~2s2p^3       $ & $^3P^e$ & $^3S^o$ &  1 (  1)&  3
(  3)& 50.3320 &  44.6040 &  7.773E-02 &  6.83E+09 \\
 $2s^22p^2      $ & $-~2s2p^3       $ & $^3P^e$ & $^1P^o$ &  1 (  1)&  3
(  4)& 50.3320 &  43.9660 &  8.405E-03 &  9.12E+08 \\
 $2s^22p^2      $ & $-~2s^22p3s     $ & $^3P^e$ & $^3P^o$ &  1 (  1)&  3
(  5)& 50.3320 &  21.6590 &  5.665E-02 &  1.25E+11 \\
 $2s^22p^2      $ & $-~2s^22p3s     $ & $^3P^e$ & $^1P^o$ &  1 (  1)&  3
(  6)& 50.3320 &  21.3810 &  3.441E-04 &  7.72E+08 \\
 $2s^22p^2      $ & $-~2s^22p3d     $ & $^3P^e$ & $^3D^o$ &  1 (  1)&  3
(  7)& 50.3320 &  19.2530 &  8.101E-01 &  2.10E+12 \\
\hline
\end{tabular}
\end{table}

\pagebreak

\begin{table}
\noindent{Table 7b. Fine structure transitions of Ar XIII are ordered in
$LS$ multiplets and are compared with previous values. $a\pm b$ means 
$a\times 10^b$; notation 'O' means for others and 'P' for present.
\\ }
\scriptsize
\begin{tabular}{llllrrllll}
\noalign{\smallskip}
\hline
\noalign{\smallskip}
\multicolumn{1}{c}{$C_i$} &\multicolumn{1}{c}{$C_j$} & $S_iL_i\pi_i$ &
$S_jL_j\pi_j$ & $g_i(I_i)$ & $g_j(I_j)$ & $f_{ij}(O)$ & $f_{ij}(P)$ &
$A_{ji}(s^{-1},O)$&$A_{ji}(s^{-1},P)$\\
 \noalign{\smallskip}
\hline
 \noalign{\smallskip}
$2s22p2$ & $ -2s2p3  $ &$^3P0$ & $^ 3D1$ & 1( 1)&3( 1)&$8.267-2^1$&7.370-2 &&\\ 
$2s22p2$ & $ -2s2p3  $ &$^3P0$ & $^ 3D1$ & 3( 1)&3( 1)&$1.163-2^1$&1.078-2 &&\\ 
$2s22p2$ & $ -2s2p3  $ &$^3P0$ & $^ 3D1$ & 3( 1)&5( 2)&$5.865-2^1$&5.352-2 &&\\ 
$2s22p2$ & $ -2s2p3  $ &$^3P0$ & $^ 3D1$ & 5( 1)&3( 2)&$1.487-4^1$&9.623-6 &&\\ 
$2s22p2$ & $ -2s2p3  $ &$^3P0$ & $^ 3D1$ & 5( 1)&5( 2)&$3.399-3^1$&2.623-3 &&\\ 
$2s22p2$ & $ -2s2p3  $ &$^3P0$ & $^ 3D1$ & 5( 1)&7( 1)&$5.037-2^1$&4.507-2 &&\\

$2s22p2$ & $ -2s2p3  $ &$^3P0$ & $^ 3D1$ & 9 & 15 &$ 5.33-2^2$& 5.209-2  &&\\
\end{tabular}
\end{table}

\clearpage
%\pagebreak

\begin{table}
\noindent{Table 8a. Sample set of reprocessed $f$- and $A$-values of
Fe XXI where observed transition energies replace the calculated ones. 
$I_i$ and $I_j$ are the level indices.
\\ }
\scriptsize
\begin{tabular}{llllcrrrcc}
\noalign{\smallskip}
\hline
\noalign{\smallskip}
\multicolumn{1}{c}{$C_i$} &\multicolumn{1}{c}{$C_j$} &
\multicolumn{1}{c}{$S_iL_i\pi_i$} & \multicolumn{1}{c}{$S_jL_j\pi_j$} &
$g_i(I_i)$ & $g_j(I_j)$ & \multicolumn{1}{c}{$E_i(Ry)$} & 
\multicolumn{1}{c}{$E_j(Ry)$} & $f$ & $A(sec^{-1})$ \\
 \noalign{\smallskip}
\hline
 \noalign{\smallskip}
 $2s^22p2     $ & $-~2s2p^3      $ & $^3P^e$ & $^3D^o$ &  1 (  1)&  3
(  1)& 124.1100 & 117.0400 &  8.491E-02 &  1.14E+10 \\
 $2s^22p^2     $ & $-~2s2p^3      $ & $^3P^e$ & $^3P^o$ & 1 (  1)&  3
(  2)& 124.1100 & 115.6900 &  2.252E-02 &  4.27E+09 \\
 $2s^22p^2     $ & $-~2s2p^3      $ & $^3P^e$ & $^3S^o$ &  1 (  1)&  3
(  3)& 124.1100 & 114.1300 &  3.579E-02 &  9.55E+09 \\
 $2s^22p^2     $ & $-~2s2p^3      $ & $^3P^e$ & $^1P^o$ &  1 (  1)&  3
(  4)& 124.1100 & 112.6200 &  1.028E-04 &  3.63E+07 \\
 $2s^22p^2     $ & $-~2s^22p3d   $ & $^3P^e$ & $^1P^o$ &  1 (  1)&  3
( 11)& 124.1100 &  48.5350 &  2.771E-02 &  4.24E+11 \\
 $2s^22p^2     $ & $-~2s^22p4d   $ & $^3P^e$ & $^3D^o$ &  1 (  1)&  3
( 39)& 124.1100 &  27.6930 &  1.027E-01 &  2.56E+12 \\
 $2s^22p^2     $ & $-~2s^22p4d   $ & $^3P^e$ & $^3P^o$ &  1 (  1)&  3
( 40)& 124.1100 &  26.7180 &  1.376E-01 &  3.49E+12 \\
\hline
\end{tabular}
\end{table}

\begin{table}
\noindent{Table 8b. Fine structure transitions of Fe XXI are ordered in 
$LS$ multiplets and are compared with previous values. $a\pm b$ means 
$a\times 10^b; notation 'O' means for others and 'P' for present$.
\\ }
\scriptsize
\begin{tabular}{llllrrllll}
\noalign{\smallskip}
\hline
\noalign{\smallskip}
\multicolumn{1}{c}{$C_i$} &\multicolumn{1}{c}{$C_j$} & $S_iL_i\pi_i$ &
$S_jL_j\pi_j$ & $g_i(I_i)$ & $g_j(I_j)$ & $f_{ij}(O)$ & $f_{ij}(P)$ &
$A_{ji}(s^{-1}$(O) & $A_{ji}(s^{-1}$(P) \\
 \noalign{\smallskip}
\hline
 \noalign{\smallskip}
$2s22p2$ & $-2s2p3$ &  $3P0$ & $^3D1$ &  1 (1)&3 (1) &$9.3-2^1$ & 8.49-2 &&\\
$2s22p2$ & $-2s2p3$ &  $3P0$ & $^3D1$ &  3 (1)&3 (1) &$2.4-3^1$ & 2.45-3 &&\\
$2s22p2$ & $-2s2p3$ &  $3P0$ & $^3D1$ &  3 (1)&5 (2) &$5.1-2^1$ & 4.62-2&&\\ 
$2s22p2$ & $-2s2p3$ &  $3P0$ & $^3D1$ &  5 (1)&3 (1) &$1.5-4^1$ & 5.02-3&&\\
$2s22p2$ & $-2s2p3$ &  $3P0$ & $^3D1$ &  5 (1)&5 (2) &$4.4-5^1$ & 7.64-6&&\\ 
$2s22p2$ & $-2s2p3$ &  $3P0$ & $^3D1$ &  5 (1)&7 (1) &$2.95-2^1$& 2.70-2&&\\ 

$2s22p2$ & $-2s2p3$ & $3P0$ & $^3D1$ & 9 & 15 &$4.5-2^1,3.71-2^2$ &4.20-2&&\\
\end{tabular}
\end{table}

\begin{table}
\noindent{Table 9. Level indices of the calculated fine structure energy 
levels for various $J\pi$ symmetries that have been observed and among 
which the allowed transitions have been reprocessed using the observed 
energies. $n_j$ is the number of observed levels in each symmetry. \\ }
\scriptsize
\begin{tabular}{lrl}
\noalign{\smallskip}
\hline
\noalign{\smallskip}
$J\pi$ & $n_j$ & level indices \\
 \noalign{\smallskip}
\hline
 \noalign{\smallskip}
 \multicolumn{3}{c}{Ar XIII} \\
 \noalign{\smallskip}
\hline
 \noalign{\smallskip}
$0^e$ & 3 & 1,2,3\\
$0^o$ & 6 & 1,2,3,8,12,18\\
$1^e$ & 2 & 1,2\\
$1^o$ &17 & 1,2,3,4,5,6,7,8,9,16,18,20,21,30,31,43,45\\
$2^e$ & 5 & 1,2,3,4,14\\
$2^o$ &15 & 1,2,3,4,5,6,7,8,9,15,16,19,29,44,46\\
$3^e$ & 1 & 6\\
$3^o$ & 5 & 1,2,3,4,8\\
$4^o$ & 1 & 1\\
 \noalign{\smallskip}
\hline
 \noalign{\smallskip}
 \multicolumn{3}{c}{Fe XXI} \\
 \noalign{\smallskip}
\hline
 \noalign{\smallskip}
$0^e$ & 4 & 1,2,3,4 \\
$0^o$ & 1 & 1 \\
$1^e$ & 2 & 1,2 \\
$1^o$ & 8 & 1,2,3,4,11,39,40,67 \\
$2^e$ & 4 & 1,2,3,4 \\
$2^o$ &11 & 1,2,3,4,6,8,9,39,40,41,72 \\
$3^o$ & 9 & 1,2,3,4,22,23,24,50,51 \\
\noalign{\smallskip}
\hline
\end{tabular}
\end{table}

\end{document}